\documentclass[epj,draft]{svjour}
\usepackage{psfig}

\begin{document}
%
\title{Localization properties of two interacting particles in a
  quasi-periodic potential with a metal-insulator transition}
\author{Andrzej Eilmes\inst{1} \and Rudolf A.\ R\"{o}mer\inst{2}
  \and Michael Schreiber\inst{2}}
\institute{Department of Computational Methods in Chemistry, Jagiellonian
  University, Ingardena 3, 30-060 Krak\'{o}w, Poland \and Institut f\"{u}r
  Physik, Technische Universit\"{a}t, D-09107 Chemnitz, Germany}
%
\date{Received: / Revised version:}
\abstract{ We study the influence of many-particle interactions on a
  metal-insulator transition. We consider the two-interacting-particle
  problem for onsite interacting particles on a one-dimensional
  quasiperiodic chain, the so-called Aubry-Andr\'{e} model. We show
  numerically by the decimation method and finite-size scaling that
  the interaction does not modify the critical parameters such as the
  transition point and the localization-length exponent. We compare
  our results to the case of finite density systems studied by means
  of the density-matrix renormalization scheme.
}
\titlerunning{Localization properties of TIP in a quasi-periodic potential with a MIT}
\maketitle


\section{Introduction}
\label{sec-intro}

The metal-insulator transition (MIT) in disordered electronic systems
has been the subject of intense research activities over the last two
decades and still continues to attract much attention. For free
electrons in disordered systems \cite{KraM93} the scaling hypothesis
of localization \cite{AbrALR79} can successfully predict many of the
universal features of the MIT. However, the influence of many-particle
interactions on the MIT is not equally well understood \cite{BelK94}
and recent investigations of an apparent MIT in two-dimensional (2D)
systems even question the main assumptions of the scaling hypothesis
\cite{KraDS96,BelK98,SimKSP97,SimKSp98,KraSSK97,PudBPB97}.
A simple theoretical approach to the interplay of interactions and
disorder is based on the two-interacting-particles (TIP) problem in 1D
random \cite{She94,She96a,RomS97a} or quasiperiodic potentials
\cite{She96b,EilGRS99}. Furthermore, numerical results for spinless
fermions at finite particle density have given additional insight
\cite{SchSSS98,SchJWP98,WSchJP01}. In general, these investigations have shown
that changes in the wave function interferences due to many-particle
interactions \cite{RomSV99,RomSV00} can lead to a rather large
enhancement of the localization lengths in 1D and 2D
\cite{SchJWP98,LeaRS99,RomLS99b}.

The standard approach for computing localization len\-gths in
disordered, non-interacting systems is the transfer-matrix method
\cite{MacK83}. It has been used for investigations of the enhancement
of the TIP localization length in a 1D random potential
\cite{RomS97a,FraMPW95} where there is no MIT as all wave functions
are always localized. Other numerical approaches to the TIP problem
have been based on the time evolution of wave packets
\cite{She94,BriGKT98}, exact diagonalization \cite{WeiMPF95} or Green
function approaches \cite{LeaRS99,OppWM96,SonK97}.

In the single-particle case, the 1D quasiperiodic Aubry-Andr\'{e}
model is known rigorously to exhibit an MIT for all states in the
spectrum as a function of the quasiperiodic potential strength $\mu$
\cite{AubA80}.  The ground state wave function is extended for $\mu <
1$ and localized for $\mu >1$. The system at $\mu_{\rm c}=1$ is
critical: there the wave functions decrease algebraically, not
exponentially as in the localized case.
Recently, we examined this model for TIP by means of the
transfer-matrix method together with a careful finite-size-scaling
analysis \cite{EilGRS99} following earlier analytical work of Refs.\ 
\cite{BarBJS96,BarBJS97}.  We showed that the model for TIP exhibits
an MIT as a function of $\mu$ at $\mu_{\rm c}=1$ as in the
single-particle case. Our finite-size-scaling results for onsite
(Hubbard) interaction suggest that the critical behavior, i.e., the
value for the critical exponent $\nu$ of the correlation length, is
also not affected by the interaction \cite{EilGRS99}. However, it has
been demonstrated \cite{RomS97a,LeaRS99} that a transfer-matrix-method
approach applied to the TIP problem without finite-size scaling leads
to unreliable localization lengths, i.e., it systematically
overestimates the TIP localization length $\lambda_2$ in finite-sized
samples in the case of vanishing interaction ($U=0$). In addition,
simple extrapolations to infinite sample size \cite{RomS97a,FraMPW95}
may lead to an underestimation of $\lambda_2$ \cite{SonO98}.
An alternative approach, which does not suffer from the above problem,
is based on the decimation method and has also been applied recently
to TIP in a 1D random potential \cite{LeaRS99}. This encouraged us to
reexamine the localization lengths for TIP in 1D quasiperiodic
potentials with Hubbard interaction with the decimation method. As we
shall show in the following, we find that the general conclusions of
Ref.\ \cite{EilGRS99} remain valid, i.e., the MIT is not affected by
the interaction. The critical properties of the single-particle
transition at $\mu_{\rm c}=1$ are not altered within the accuracy of
our calculation.  One-parameter scaling is obeyed for onsite
interaction strengths up to $U=10$.

As an independent extension of these low-density results, Chaves and
Satija \cite{ChaS97} have studied a model of nearest-neighbor
interacting spinless fermions \cite{YanY66a} at finite particle density in
the same quasiperiodic potential by means of Lanczos diagonalization for
small systems up to chain size $M=13$. They have found evidence for a critical
region. In order to reach much larger system sizes for interacting
systems, one can employ the numerical density-matrix renormalization group
(DMRG) \cite{Whi93}. With the DMRG the ground state properties in 1D
can be obtained very accurately \cite{PaiPR97,Sch99}. In a recent
paper \cite{SchRS01}, we studied the quasiperiodic model of Ref.\ 
\cite{ChaS97} at various densities and interaction strengths $V$ by
DMRG. We compare the results with the present TIP data at the end of the
paper.

The paper is organized as follows. In section \ref{sec-tip} we
describe the Hamiltonian of our TIP system and explain how to obtain
the TIP localization lengths via decimation method. In section
\ref{sec-tip-results}, we comment on the particular
finite-size-scaling method employed and present the estimated critical
parameters. We summarize and conclude in section \ref{sec-concl}.

\section{The TIP system and the numerical approach}
\label{sec-tip}

The Hamiltonian for TIP in the 1D quasiperiodic potential of the
Aubry-Andr\'{e} model is given as
\begin{eqnarray}
  {\bf H} & = & \sum_{n,m} \vert n, m \rangle\langle n+1,m\vert
    + \vert n, m \rangle\langle n,m+1\vert + {\rm h.\ c.} \nonumber \\
  & & \mbox{ } + \vert n, m \rangle \left[ \mu_n + \mu_m +
    U(n,m) \right] \langle n,m\vert\ .
\label{eq-ham}
\end{eqnarray}
Here $\mu_m \equiv 2 \mu \cos( \alpha m + \beta)$ is the quasiperiodic
potential of strength $\mu$ with $\alpha /2 \pi$ being an irrational
number.  $\beta$ is an arbitrary phase shift and we choose $\alpha/2
\pi= (\sqrt{5} -1)/2$, i.e., the inverse of the golden mean. This
value of $\alpha/2 \pi$ may be approximated by the ratio of successive
Fibonacci numbers --- $F_n=F_{n-2}+F_{n-1}= 0$, $1$, $2$, $3$, $5$,
$8$, $13$, $\ldots$ --- as is customary in the context of
quasiperiodic systems \cite{Gri99}. The Hubbard onsite interaction
matrix $U(n,m)$ is diagonal, i.e., $U(n,m)=U \delta_{nm}$. The indices
$n$ and $m$ correspond to the positions of each particle on a chain of
length $M$. 
Now we use the decimation method \cite{LeaRS99,LamW80} to construct an
effective Hamiltonian for the diagonal of the $M \times M$ lattice
along which the cigar-shaped TIP wave function has its largest extent
\cite{WeiMPF95,SonO98}. The quantity of interest is the TIP
localization length $\lambda_2$ defined by the TIP Green function
${\bf G_2}(E)$ \cite{OppWM96}:
\begin{equation}
  {1\over\lambda_2} = - {1\over\vert M-1\vert} \ln\vert\langle
  1,1\vert {\bf G_2}\vert M,M\rangle\vert.
\label{eq-lambda2}
\end{equation}
For TIP in 1D and 2D random potentials, this approach has led to high
precision results supporting the proposed increase of the TIP
localization lengths due to the repulsive interaction
\cite{LeaRS99,RomLS99b}. We remark that similar data have also been
obtained for nearest-neighbor \cite{OppWM96} and long-ranged
interactions \cite{BriGKT98}.

The correlation length $\xi_\infty$ for the infinite system may be
obtained from the localization lengths $\lambda(M)$ for finite system
sizes by the using one-parameter scaling hypothesis
$\Lambda_M=f(M/\xi_\infty)$ \cite{Tho74} for the reduced localization
lengths $\Lambda_M=\lambda(M)/M$. The MIT is characterized by a
divergent correlation length $\xi_\infty(\mu)\propto |\mu-\mu_{\rm
  c}|^{-\nu}$ \cite{KraM93}.
In order to reliably extract the critical parameters from the
calculated values of $\lambda_2(M)$ one may apply a
finite-size-scaling procedure \cite{MacK83} that numerically minimizes
deviations of the data from the common scaling curve $f$. The critical
exponent $\nu$ can then be extracted by fitting the $\xi_\infty$
obtained from finite-size scaling \cite{cm0106005,cm0106006}. This
method was used previously \cite{EilGRS99} for finding the critical
parameters of the present model.

Higher accuracy can be achieved by a method applied recently
\cite{cm0106005,cm0106006,SleO99a,MilRSU00} to the MIT in the Anderson model of
localization.  We construct a family of fit functions which include
corrections to scaling such as (i) nonlinearities of the dependence of
the scaling variable on the quasiperiodic potential strength and (ii)
an irrelevant scaling variable which accounts for a shift of the
crossing point of the $\Lambda_M(\mu)$ curves as a function of $\mu$,
i.e.,
\begin{equation}
  \label{eq-Slevin}
  \Lambda_M=\tilde{f}(\chi_{\rm r} M^{1/\nu}, \chi_{\rm i} M^{y})
  \quad .
\end{equation}
where $\chi_{\rm r}$ and $\chi_{\rm i}$ are the relevant and
irrelevant scaling variables, respectively. $\tilde{f}$ is then Taylor
expanded up to order $n_{\rm i}$ in terms of the second argument
\begin{equation}
  \label{eq-Slevin2}
  \Lambda_M=\sum_{n=0}^{n_{\rm i}} \chi_{\rm i}^n M^{n
    y}\tilde{f}_n(\chi_{\rm r} M^{1/\nu}) \quad ,
\end{equation}
and each $\tilde{f}_n$ is Taylor expanded up to order $n_{\rm r}$:
\begin{equation}
  \tilde{f}_n=\sum_{i=0}^{n_{\rm r}} a_{ni} \chi_{\rm r}^i M^{i/\nu}
  \quad .
\end{equation}
Nonlinearities are taken into account by expanding $\chi_{\rm r}$ and
$\chi_{\rm i}$ in terms of $u=(\mu_{\rm c}-\mu)/\mu_{\rm c}$ up to
order $m_{\rm r}$ and $m_{\rm i}$, respectively,
\begin{equation}
  \label{eq-Slevin-Var}
  \chi_{\rm r}(u)=\sum_{n=1}^{m_{\rm r}} b_n u^n , \quad \chi_{\rm
    i}(u)=\sum_{n=0}^{m_{\rm i}} c_n u^n \quad ,
\end{equation}
with $b_1=c_0=1$. The fit function is being adjusted to the data by
choosing the orders $n_{\rm i}, n_{\rm r}, m_{\rm r}, m_{\rm i}$ up to
which the expansions are carried out.  Of course, the orders have to
be taken not too large to keep the number of fit parameters $a_{ni}$,
$b_n$, and $c_n$ reasonably small.

\section{Numerical results for TIP}
\label{sec-tip-results}

We calculate $\lambda_2$ at energy $E=0$ for $20$ values of the
Hubbard interaction, i.e., $U=0$ (the non-interacting single-particle
case), $0.1$, $\ldots$, $0.9$, $1$, $2$, $\ldots$, $10$ for $6$ system
sizes $M = 13$, $21$, $34$, $55$, $89$, $144$. For $U=0$ and $1$, we
also have data for $M=233$ and $377$. The quasiperiodic potential
strengths $\mu$ is chosen close to the single-particle transition at
$\mu_{\rm c}\approx1$ and ranges typically from $0.95$ to $1.05$. As
in Ref.\ \cite{EilGRS99} we average the results over different
randomly chosen phase shifts $\beta$ in order to reduce the
fluctuations. The number of $\beta$ values used in this averaging
ranges from $5000$ for $M=13$ to $1000$ for $M=377$. In order to
perform the non-linear fit necessary for the finite-size-scaling
procedure as outlined in section \ref{sec-tip}, we use the
Levenberg-Marquardt method \cite{MilRSU00,PreFTV92}.  As the
decimation-method data --- like the transfer-matrix-method results
\cite{EilGRS99} --- are still rather noisy we have to suitably limit
the ranges of the quasiperiodic potential strength $\mu$ and/or the
system sizes $M$ used for fitting the data.

For $U=0$ and $1$, which were examined by the transfer-matrix method
in detail \cite{EilGRS99}, the best fit is obtained for $n_r=3$,
$n_i=2$, $m_r=2$ and $m_i=1$. For $U=0$ we use the data for $\mu$
ranging from $0.96$ to $1.01$ and $M=55,89,144,233$, and $377$; for
$U=1$ we use all system sizes $M=13$, $\ldots$, $377$ and $0.97 \le
\mu \le 1.05$.
Figures  \ref{fig-dat-u0} and \ref{fig-dat-u1} show the resulting TIP
localization lengths for $U=0$ and $1$. Also shown are the fits of the
finite-size-scaling curves to the data as given by Eq.\
(\ref{eq-Slevin}) for $U=0$ and $1$, respectively. We find that for
both $U$ values, there is an apparent transition close to $\mu_{\rm
  c}=1$. For the case $U=0$, we also observe a systematic shift of the
crossing point with increasing system sizes necessitating the
inclusion of an irrelevant scaling variable as discussed in section
\ref{sec-tip}. The transition point is not so clearly distinguished
for $U=1$, albeit the different behavior for $\mu \le 1$ and $\mu
\ge 1$, namely the increase and decrease, respectively, of
$\Lambda_M$ with increasing $M$, is clearly seen.

The corresponding plots of the scaling curves are displayed in the
Figs. \ \ref{fig-fss-u0} and \ref{fig-fss-u1}. The scaling
curves are much better than reported previously \cite{EilGRS99} for
the transfer-matrix-method data. The critical parameters can
consequently be estimated to be $\mu_{\rm c}=0.989 \pm 0.001$,
$\nu=1.00 \pm 0.15$ for $U=0$ and $\mu_{\rm c}=0.997 \pm 0.001$,
$\nu=1.19 \pm 0.16$ for $U=1$. The irrelevant scaling exponents are
close to $y=1.8 \pm 0.2$ and $y=0.15 \pm 0.1$ for $U=0$ and $1$,
respectively. Note that the quoted errors correspond to the standard
deviations estimated from the non-linear fit procedure. In this way
the accuracy is significantly overestimated. Since it is apriori not
clear, which values $n_{\rm i}, n_{\rm r}, m_{\rm r}, m_{\rm i}$ to
use, we estimate the true errors from a comparison of various fits
with different $n_{\rm i}, n_{\rm r}, m_{\rm r}, m_{\rm i}$. Even in
the case of extremely high precision data close to the MIT in the
Anderson model of localization, this has been shown \cite{MilRSU00} to
increase the error by one order of magnitude. Therefore we conclude
that the interaction strength $U$ for TIP does not influence the MIT
in the quasiperiodic potential within the accuracy of the present
calculation.

Further results for larger $U$ values are collected in Table
\ref{tab-U}. The expansion orders $n_{\rm i}, n_{\rm r}, m_{\rm r},
m_{\rm i}$, the system sizes and ranges of the quasiperiodic potential
strength have been chosen in order to minimize the $\chi^2$ statistics
and to get the most convincing scaling fit. Furthermore, one has to check
that various initial parameters ($a_n, b_n, c_n$) converge to the same
values of the critical quasiperiodic potential strength $\mu_{\rm c}$ and the
critical exponent $\nu$.
Figures \ref{fig-tip-mucrit} and \ref{fig-tip-nucrit} show the values
obtained in this way. For almost all cases the critical quasiperiodic
potential strength $\mu_{\rm c}$ remains close to $1$, the only exceptions
are $U=0$ and $0.1$, when $\mu_{\rm c}=0.99$ and $0.98$, respectively.
However, since we know that the transition in the single-particle case
is exactly at $\mu_{\rm c}=1$ \cite{AubA80}, this observation can be used
to estimate the true error of the estimate for $\mu_{\rm c}$. Thus
comparing with the $\mu_{\rm c}$ estimates for $U\neq 0$, we find that
the errors calculated within the non-linear fitting procedure are
significantly underestimated as discussed above. We therefore conclude
that within the accuracy of our
calculation there is no change of the critical quasiperiodic potential
strength $\mu_{\rm c}$ for the Hubbard interaction in the range
$0\leq U \leq 10$. The same argument leads to the conclusion that within the
error bars the critical exponent $\nu$ does not change with the
Hubbard interaction strength and is close to $1$. This is an agreement
with the previous results obtained by the transfer-matrix method and
finite-size scaling \cite{EilGRS99}. We stress that the critical
exponents can only be obtained with much less accuracy than the transition
point $\mu_{\rm c}$ as shown in Table \ref{tab-U}.

\section{Conclusions}
\label{sec-concl}

In this work, we have studied the interplay of disorder
and interactions for a quantum system at very low density (TIP).
We calculated the pair localization lengths for a quasiperiodic potential
and Hubbard interaction by means of the decimation method and extracted
the critical parameters from the fit using the one-parameter scaling
hypothesis. For both non-interacting particles as well as onsite interaction
we obtain the value of the critical quasiperiodic potential strength
$\mu_{\rm c}=1$ and the critical exponent $\nu\approx 1$ in agreement
with the previous results of transfer-matrix-method calculations and
finite-size scaling \cite{EilGRS99}.  The results for $U > 1$ show that
this conclusion remains valid also for much stronger interactions.

Let us briefly compare these results to the finite density situation.
For $N$ interacting spinless fermions on a 1D ring of circumference $M$
with Aubry-Andr\'{e} onsite potential $\mu$ and nearest-neighbour
interaction $V$ it is possible to treat system lengths up
to about $M\approx 100-200$ using the DMRG. We applied \cite{SchRS01}
the finite lattice algorithm for non-reflectionsymmetric models as
described in \cite{Sch96}.
For a system of free fermions at finite density like $\rho=1/2$
(incommensurate compared to the wave vector of the quasiperiodic potential ---
an irrational multiple of $\pi$), we reproduced \cite{SchRS01} the expected
transition at $\mu_{\rm c}=1$ in agreement with Refs.\ \cite{EilGRS99,ChaS97}.
For attractive and repulsive interactions at $\rho=1/2$ the numerial
results are available for only two system sizes ($M=34$ and $144$),
therefore conclusions about these regimes appear rather speculative.
At commensurate densities $\rho_i \approx \lim_{n\rightarrow\infty}
F_{n-i}/F_n \approx 0.618$, $0.382$, $0.236$, and $0.146$ ---
corresponding to $i=1$, $\ldots$, $4$ --- and in the repulsive regime
(nearest-neighbour interaction $V>0$), the ground state is localized for
$\mu>0$ \cite{SchRS01} in agreement with previous studies for disordered
and periodically disturbed systems \cite{SchSSS98,Sch99}. The above increase of the
localization lengths as predicted by the arguments for TIP
\cite{She94} is most likely too small \cite{SchJWP98} to be detected
by the present accuracy.
For attractive interactions $V$, all densities $\rho_i$ and $\mu\to 0$, the
system shows a {\em Peierls-like} transition from
insulating to metallic phase at $V\approx -1.4$ \cite{SchRS01} in
agreement with the weak-coupling renormalization group treatment
\cite{VidMG99} of spinless fermions on a Fibonacci lattice.

In conclusion, we have studied the influence of interactions on an MIT in
a quasiperiodic model in 1D. Our results suggest that the
delocalization found for low density TIP in the localized
phase cannot simply be extrapolated to the finite-density situation.
At finite densities, other effects such as a Peierls-like
commensurability become important and dominate the transport
properties.

\begin{acknowledgement}
  We thank M.\ Leadbeater for help with the decimation method and C.\ 
  Schuster for stimulating discussions.  We gratefully acknowledge the
  support of the SMWK and the Deutsche For\-schungs\-gemein\-schaft
  within Sonderforschungsbereich~393.
\end{acknowledgement}
%
%


\clearpage

%
%

\onecolumn
\begin{table}
\caption{
  Values of the critical quasiperiodic disorder strength $\mu_{\rm c}$
  and the critical exponent $\nu$ obtained by the non-linear fit for
  various $U$ values.  The first row for each $U$ gives values and the
  orders $n_{\rm i}$, $m_{\rm i}$, used in the expansion
  (\ref{eq-Slevin2}--\ref{eq-Slevin-Var}), for which the best fits
  have been obtained. In all cases we find $n_{\rm r}=3$ and $m_{\rm
    r}=2$.  For $\mu$ and $M$ the range of the values which were used
  in the fit is given. The second row contains values of the critical
  parameters obtained from a weighted average of fits for various
  choices of $n_{\rm i}$ and $m_{\rm i}$.  }
\label{tab-U}
\begin{tabular}{|c|c|c|r|r|l@{$\pm$}l|l@{$\pm$}l|}
\hline\noalign{\smallskip}
$U$  & $\mu$
     & $M$
     & $n_{\rm i}$ &  $m_{\rm i}$
     & \multicolumn{2}{c}{$\mu_{\rm c}$}
     & \multicolumn{2}{c}{$\nu$}\\
\noalign{\smallskip}\hline\noalign{\smallskip}
0 & $0.96-1.01$ &  $55-377$  & $2$&$1$      & $0.989$ & $0.001$ & $1.00$ & $0.15$\\
  & $0.95-1.05$ &  $13-377$  & $0-2$&$0-1$  & $0.99$  & $0.02$  & $1.3$   & $0.5$\\ 

1 & $0.97-1.05$ &  $13-377$  & $2$&$1$      & $0.997$ & $0.001$ & $1.19$ & $0.16$\\
  & $0.95-1.05$ &  $13-377$  & $0-2$&$0-1$  & $0.99$  & $0.01$  & $1.3$  & $0.4$\\

2 & $0.97-1.05$ &  $55-144$  & $0$&$0$      & $1.001$ & $0.002$ & $1.14$ & $0.11$\\
  & $0.95-1.05$ &  $13-144$  & $0-2$&$0-1$  & $0.99$  & $0.02$  & $1.5$  & $1$\\

3 & $0.95-1.05$ &  $13-144$  & $2$&$1$      & $1.000$ & $0.002$ & $1.16$ & $0.08$\\
  & $0.95-1.05$ &  $13-144$  & $0-2$&$0-1$  & $1.00$  & $0.02$  & $1.8$  & $1$\\

4 & $0.97-1.05$ &  $55-144$  & $0$&$0$      & $1.000$ & $0.003$ & $1.12$ & $0.10$\\
  & $0.95-1.05$ &  $13-144$  & $0-2$&$0-1$  & $1.00$  & $0.01$  & $1.5$  & $0.8$\\

5 & $0.95-1.05$ &  $13-144$  & $1$&$1$      & $1.002$ & $0.002$ & $1.20$ & $0.09$\\
  & $0.95-1.05$ &  $13-144$  & $0-2$&$0-1$  & $1.00$  & $0.01$  & $1.2$  & $0.3$\\

6 & $0.95-1.05$ &  $55-144$  & $0$&$0$      & $0.999$ & $0.002$ & $1.28$ & $0.08$\\
  & $0.95-1.05$ &  $13-144$  & $0-2$&$0-1$  & $1.00$  & $0.02$  & $1.3$  & $0.1$\\

7 & $0.95-1.05$ &  $55-144$  & $0$&$0$      & $0.997$ & $0.002$ & $1.28$ & $0.07$\\
  & $0.95-1.05$ &  $13-144$  & $0-2$&$0-1$  & $1.00$  & $0.01$  & $1.5$  & $0.6$\\

8 & $0.97-1.05$ &  $55-144$  & $0$&$0$      & $1.001$ & $0.002$ & $1.16$ & $0.08$\\
  & $0.95-1.05$ &  $13-144$  & $0-2$&$0-1$  & $0.99$  & $0.02$  & $1.4$  & $0.4$\\

9 & $0.97-1.05$ &  $13-144$  & $1$&$1$      & $1.000$ & $0.001$ & $1.15$ & $0.05$\\
  & $0.95-1.05$ &  $13-144$  & $0-2$&$0-1$  & $1.00$  & $0.01$  & $1.4$  & $0.5$\\

10& $0.97-1.05$ &  $55-144$  & $0$&$0$      & $1.000$ & $0.002$ & $1.23$ & $0.08$\\
  & $0.95-1.05$ &  $13-144$  & $0-2$&$0-1$  & $1.00$  & $0.01$  & $1.4$  & $0.4$\\
\noalign{\smallskip}\hline
\end{tabular}
\end{table}
%
%
\twocolumn
\newcommand{\figwidth}{0.95\columnwidth}

\vspace*{10ex}
\begin{figure}
  \centerline{\psfig{figure=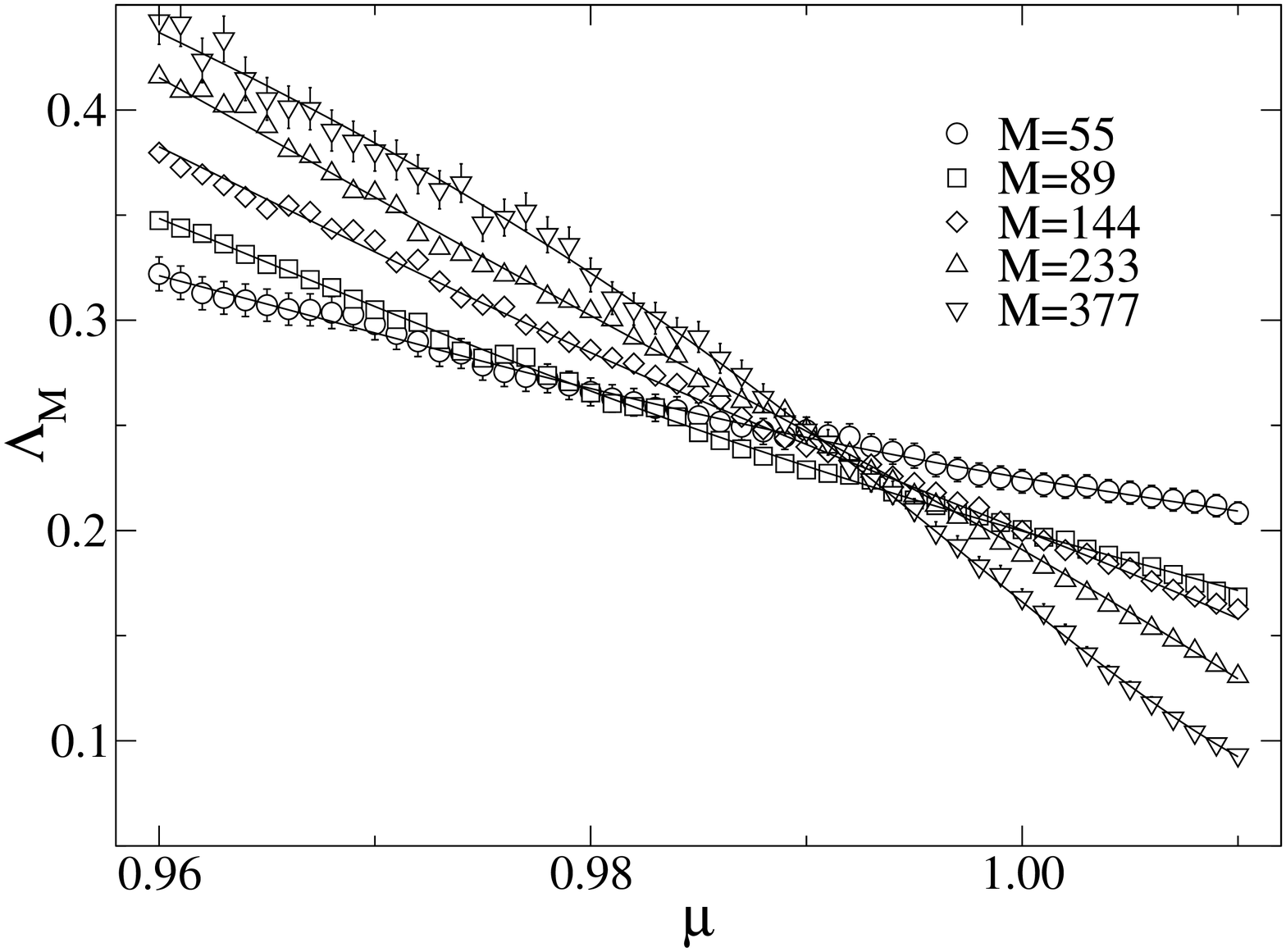,width=\figwidth}}
  \caption{
    Reduced localization lengths $\Lambda_M$ versus quasiperiodic
    disorder strength $\mu$ for $U=0$. For clarity, only error bars
    for $M=55$ and $377$ are given. The lines are the fits to the data
    given by Eq.\ (\protect\ref{eq-Slevin}).}
\label{fig-dat-u0}
\end{figure}

\vspace*{10ex}
\begin{figure}
  \centerline{\psfig{figure=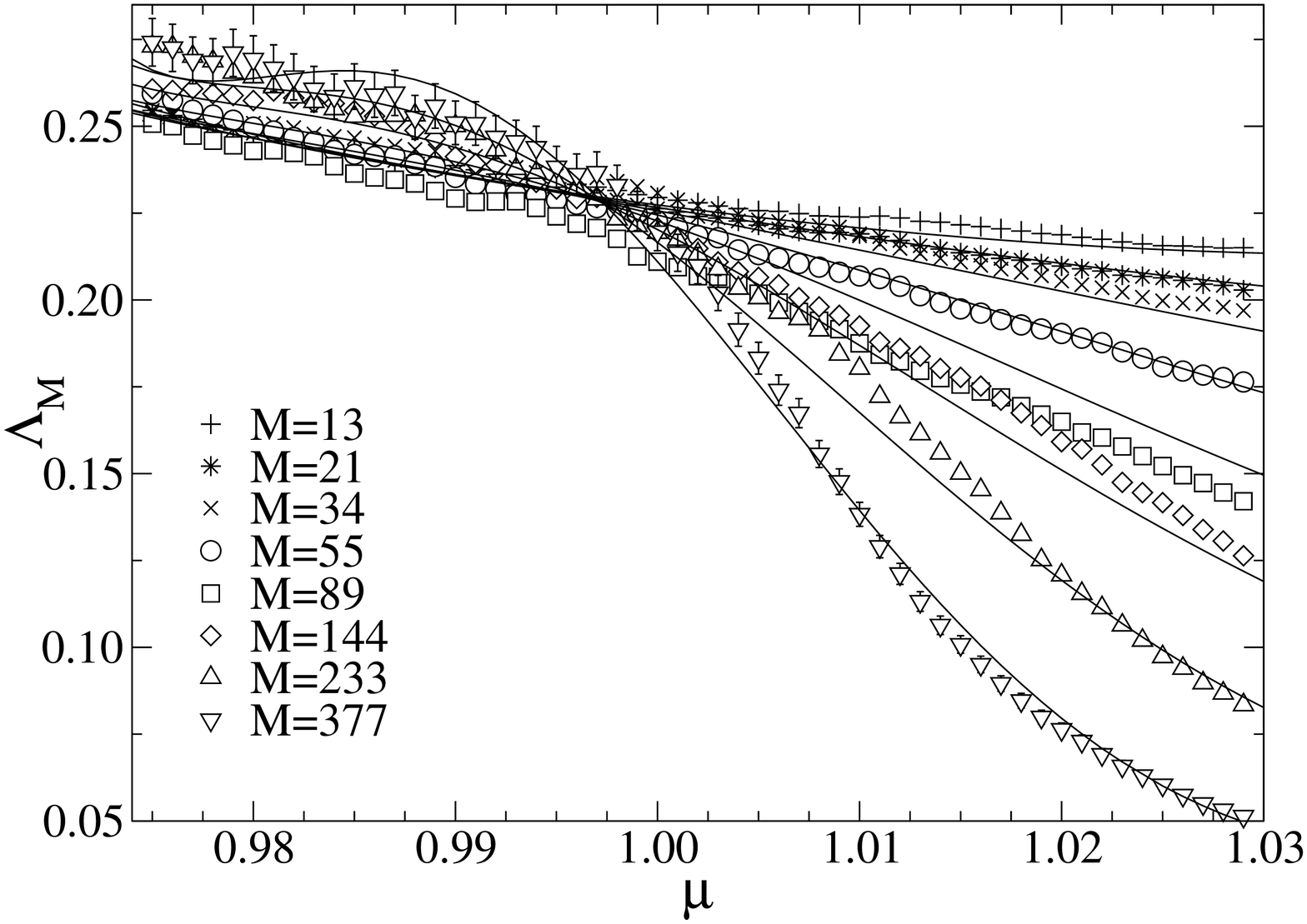,width=\figwidth}}
  \caption{
    Reduced localization lengths $\Lambda_M$ versus quasiperiodic
    disorder strength $\mu$ for $U=1$. For clarity, only error bars
    for $M=377$ are given. The lines are the fits to the data given by
    Eq.\ (\protect\ref{eq-Slevin}).}
\label{fig-dat-u1}
\end{figure}

\vspace*{10ex}
\begin{figure}
  \centerline{\psfig{figure=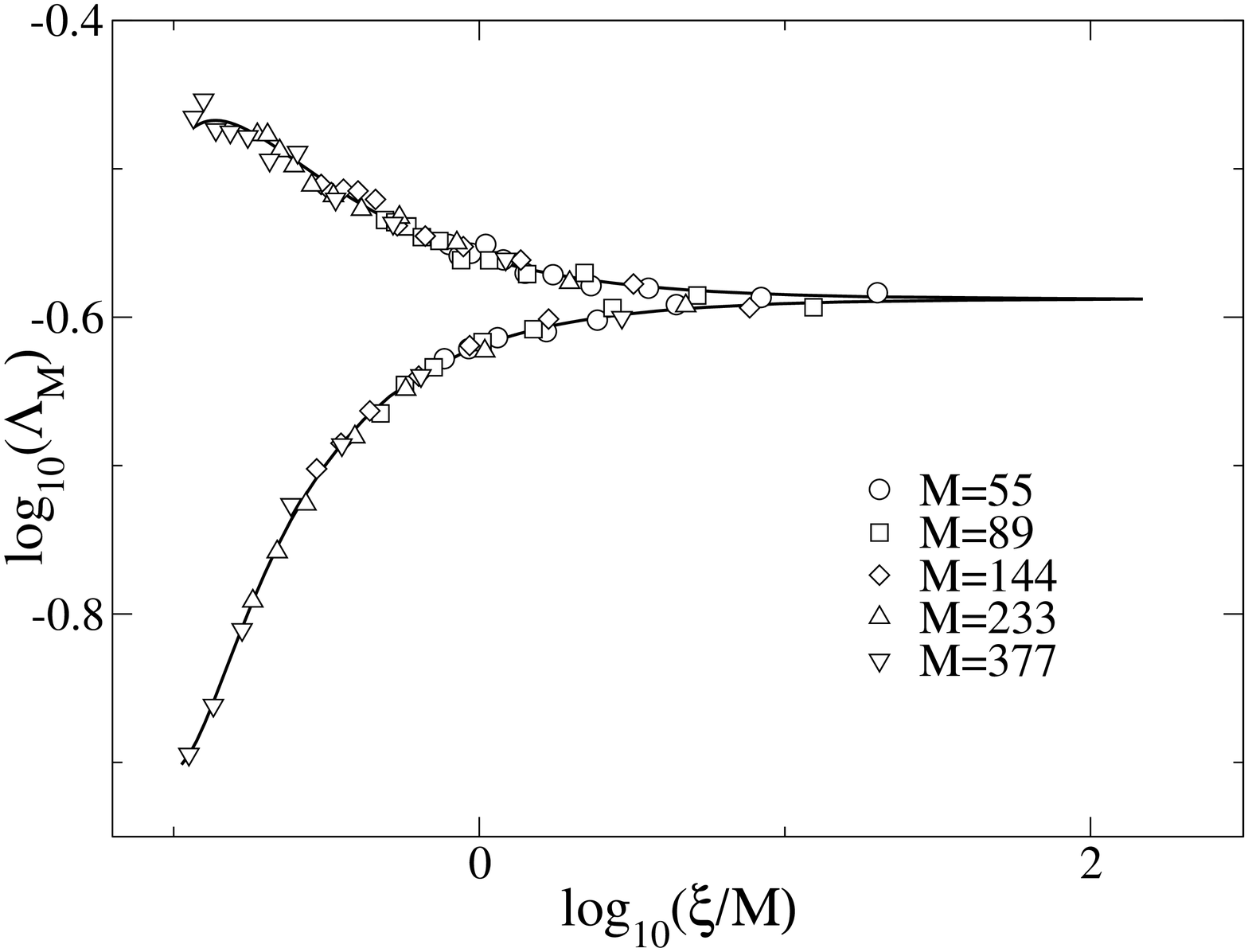,width=\figwidth}}
  \caption{
    Scaling function (solid line) and scaled data points for $U=0$.
    For clarity only every 3rd data point is represented by a symbol.}
\label{fig-fss-u0}
\end{figure}

 \vspace*{10ex}
\begin{figure}
  \centerline{\psfig{figure=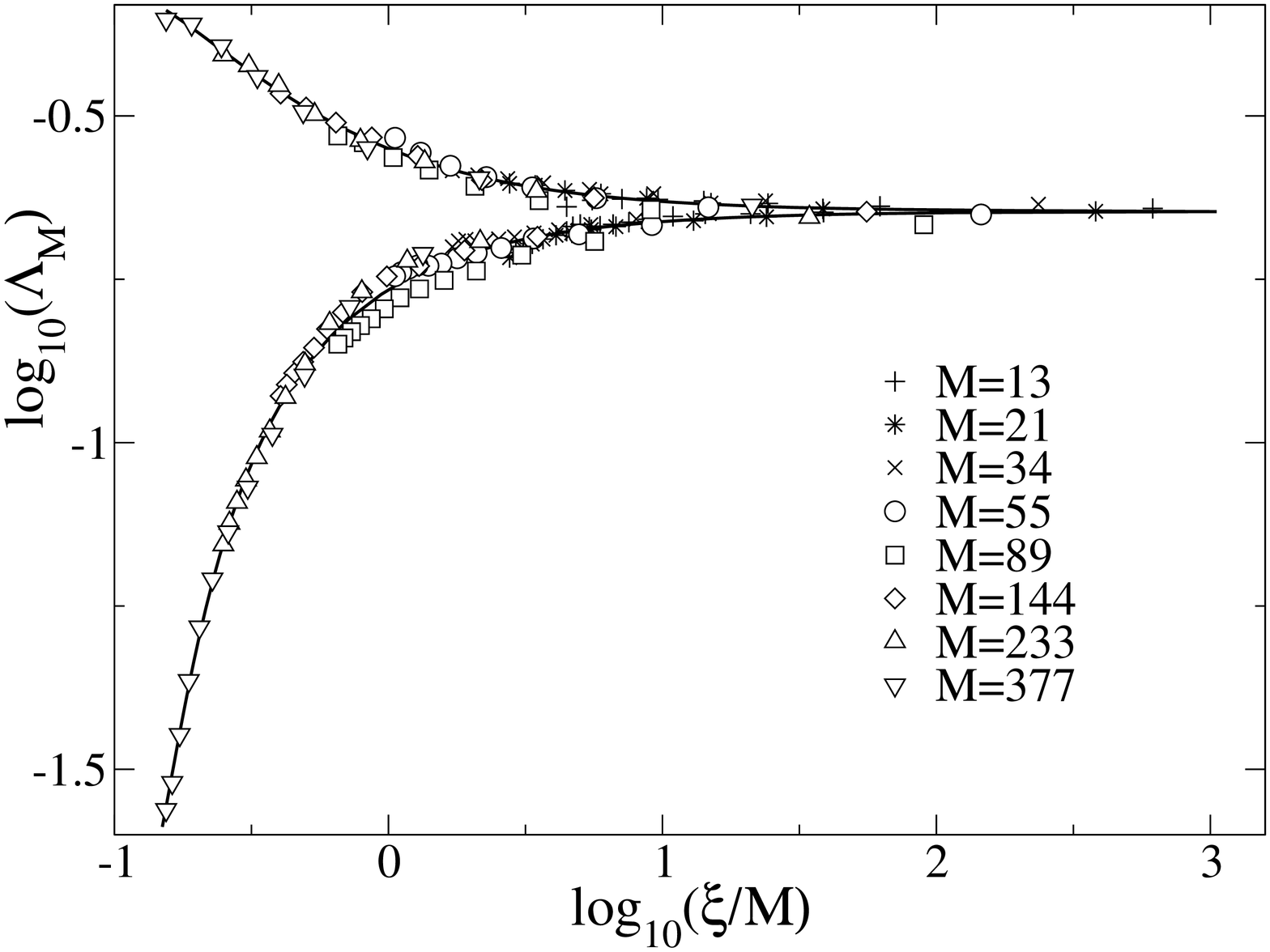,width=\figwidth}}
  \caption{
    Scaling function (solid line) and scaled data points for $U=1$.
    For clarity only every 3rd data point is represented by a symbol.}
\label{fig-fss-u1}
\end{figure}


\vspace*{10ex}
\begin{figure}
  \centerline{\psfig{figure=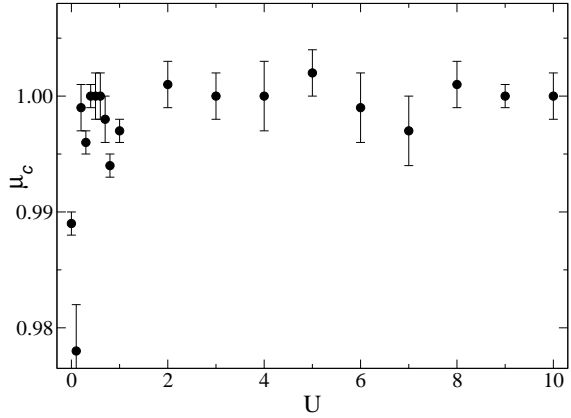,width=\figwidth}}
  \caption{
    The critical quasiperiodic potential strength $\mu_{\rm c}$ versus
    Hubbard interaction strength $U$.  Error bars mark the errors
    resulting from the Levenberg-Marquardt method of the non-linear
    fit. }
\label{fig-tip-mucrit}
\end{figure}


\vspace*{2ex}
\begin{figure}
  \centerline{\psfig{figure=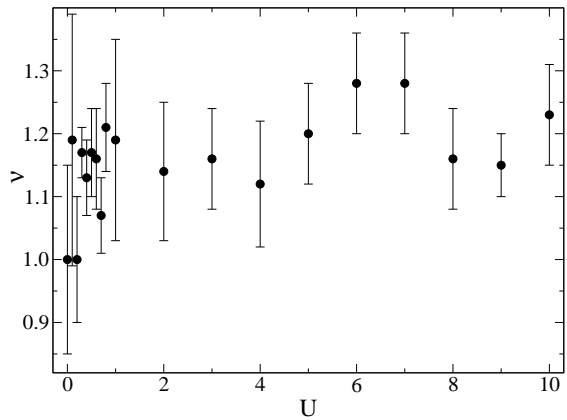,width=\figwidth}}
  \caption{
    The critical exponent $\nu$ versus Hubbard interaction strength $U$.
    Error bars mark the errors resulting from the Levenberg-Marquardt
    method of the non-linear fit. }
\label{fig-tip-nucrit}
\end{figure}


\end{document}